\documentclass[12pt]{iopart}
\usepackage{iopams,graphicx}
\begin{document}

\title[Neutrino spin-flavour conversion and mass hierarchy]{A
comprehensive study of neutrino spin-flavour conversion in supernovae and
the neutrino mass hierarchy}
\author{Shin'ichiro Ando\dag\ and Katsuhiko Sato\dag\ddag}

\address{\dag\  Department of Physics, School of Science, The University
of Tokyo, 7-3-1 Hongo, Bunkyo-ku, Tokyo 113-0033, Japan}

\address{\ddag\  Research Center for the Early Universe, School of
Science, The University of Tokyo, 7-3-1 Hongo, Bunkyo-ku, Tokyo
113-0033, Japan}

\ead{ando@utap.phys.s.u-tokyo.ac.jp}

\begin{abstract}
Resonant spin-flavour (RSF) conversions of supernova neutrinos, which is
 induced by the interaction between the nonzero neutrino magnetic moment
 and supernova magnetic fields, are studied for both normal and inverted
 mass hierarchy.
As the case for the pure matter-induced neutrino oscillation
 (Mikheyev--Smirnov--Wolfenstein (MSW) effect), we find that the RSF
 transitions are strongly dependent on the neutrino mass hierarchy as
 well as the value of $\theta_{13}$.
Flavour conversions are solved numerically for various neutrino parameter
 sets, with presupernova profile calculated by Woosley and Weaver.
In particular, it is very interesting that the RSF-induced
 $\nu_\rme\to\bar\nu_\rme$ transition occurs, if the following
 conditions are all satisfied: the value of $\mu_\nu B$ ($\mu_\nu$ is
 the neutrino magnetic moment, and $B$ is the magnetic field strength)
 is sufficiently strong, the neutrino mass hierarchy is inverted, and
 the value of $\theta_{13}$ is large enough to induce adiabatic MSW
 resonance.
In this case, the strong peak due to original $\nu_\rme$ emitted from
 neutronization burst would exist in time profile of the neutrino events
 detected at the Super-Kamiokande detector.
If this peak were observed in reality, it would provide fruitful
 information on the neutrino properties.
On the other hand, characters of the neutrino spectra are also different
 between the neutrino models, but we find that there remains degeneracy
 among several models.
Dependence on presupernova models is also discussed.
\end{abstract}

\pacs{95.85.Ry, 13.40.Em, 14.60.Pq, 97.60.Bw}


\maketitle

\section{Introduction}
\label{sec:Introduction}

A core-collapse supernova explosion is one of the most spectacular
events in astrophysics, and it attracts a great deal of attention from
many physicists and astronomers.
It also produces a number of neutrinos and 99\% of its gravitational
binding energy is transformed to neutrinos.
Therefore, neutrinos play an essential role in supernovae, and their
detection by ground-based large water \v{C}erenkov detectors, such as
Super-Kamiokande (SK) and Sudbury Neutrino Observatory (SNO), would
provide valuable information on the nature of neutrinos as well as
supernova physics.
What we can learn from the next galactic supernova has been considered
in many articles (for a review, see \cite{Raffelt02}).
For example, we can constrain the properties of neutrino oscillations,
such as the mixing angle between the first and third mass eigenstates
($\theta_{13}$), and the mass hierarchy [normal ($m_1\ll m_3$) or
inverted ($m_1\gg m_3$)] \cite{Dighe00,Takahashi03a}.

In addition to the nonzero neutrino masses and mixing angles, the
nonzero magnetic moment is another nature of neutrinos beyond the
standard model of particle physics.
If neutrinos have a nonzero magnetic moment, it leads to precession
between left- and right-handed neutrinos in sufficiently strong magnetic
fields \cite{Cisneros70,Fujikawa80}.
In general, nondiagonal elements of the magnetic moment matrix are
possible and neutrinos can be changed into different flavours and
chiralities \cite{Schechter81}.
Furthermore, with the additional effect of coherent forward scattering
by matter, neutrinos can be resonantly converted into those with
different chiralities \cite{Lim88} by a mechanism similar to the
well-known Mikheyev--Smirnov--Wolfenstein (MSW) effect
\cite{Wolfenstein78}.
This resonant spin-flavour (RSF) conversion induced by the neutrino
magnetic moment in strong magnetic fields was first introduced to
solve the solar neutrino problem, and actually gave the best fit
solution before the KamLAND result \cite{Barranco02}.
However, the recent KamLAND experiment \cite{Eguchi03} has shown that
the large mixing angle (LMA) MSW solution is the most favourable one; the
RSF mechanism is suppressed at the subdominant level.
From the KamLAND negative results for the solar antineutrino search, an
upper bound on the neutrino magnetic moment is obtained,
$\mu_\nu\lesssim 1 \times 10^{-12}\mu_{\rm B}$, where $\mu_{\rm B}$ is
the Bohr magneton \cite{Torrente-Lujan03}.
This upper bound is comparable to the most stringent limit from the
stellar cooling argument, $\mu_\nu\lesssim$(1--4)$\times
10^{-12}\mu_{\rm B}$ \cite{Ayala99}.

Although the RSF mechanism does not work at a dominant level in the
Sun, it may occur efficiently in a denser environment with stronger
magnetic field, which is actually expected in the case of core-collapse
supernovae.
The RSF conversion mechanism in supernovae has been investigated by many
authors \cite{Lim88,Voloshin88,Ando03b,Ando03d}.
Among them, Ando and Sato \cite{Ando03b} have studied the RSF effect
using a three-flavour formulation with the latest oscillation parameters,
and pointed out that the combination of the MSW and RSF effects makes
the crossing scheme very interesting to investigate.
Since the RSF conversions are very sensitive to the value of $Y_{\rm
e}$, which is the electron number fraction per nucleon, they have also
investigated the dependence of the RSF effect on presupernova models
with solar and zero metallicities \cite{Ando03d}.
It is concluded that the efficient (either complete or incomplete,
depending on presupernova models) RSF conversions occur when the
supernova magnetic field is sufficiently strong, $\mu_\nu B_0\simeq
(10^{-12}\mu_{\rm B})(10^{10}~{\rm G})$, where $B_0$ is the strength of
the magnetic field at the surface of the iron core.

However, all the past studies of three-flavour RSF effect were based on
the assumption that the neutrino mass hierarchy is normal ($m_1\ll
m_3$), although the inverted mass hierarchy ($m_1\gg m_3$) has not been
excluded at all.
For the pure MSW effect, it is well-known that the supernova neutrino
signal with the case of inverted hierarchy would be very different from
that with normal hierarchy \cite{Dighe00,Takahashi03a}.
From the analogy of the conversion mechanisms between the MSW and RSF
effects, it is easily inferred that the RSF conversions will be also
very sensitive to the mass hierarchy.
Therefore, in this paper, we study three-flavour RSF conversions with the
inverted mass hierarchy using the latest neutrino mixing parameters, and
give a comprehensive discussion concerning the dependence on the mass
hierarchy.
Since the RSF conversions are extremely sensitive to the presupernova
models as noted above, we first adopt the model of 15$M_\odot$ with
solar metallicity by Woosley and Weaver \cite{Woosley95} as a reference
model; the results with the model by Woosley \etal \cite{Woosley02} with
both solar and zero metallicity are addressed later.
In particular, we show that the RSF conversion in the case of the
inverted hierarchy with the large $\theta_{13}$ causes very different
neutrino signal from the other models, i.e., the appearance of sharp
peak of the neutronization burst in the $\bar\nu_\rme$ time profile.
If this case were realized actually, it would be clearly confirmed by
not only the neutrino spectrum but also the luminosity curve.

After the completion of our calculation, a paper appeared in which
three-flavour RSF effect was studied with both normal and inverted mass
hierarchy \cite{Ahriche03}.
The authors studied rather qualitatively, but they did not obtain the
result that the neutronization peak in the $\bar\nu_\rme$ signal
would be detected, when the efficient RSF conversion takes place in the
case of the inverted mass hierarchy and large $\theta_{13}$; our
numerical approach as well as the qualitative discussion clearly
indicate that result.
Thus, we stress that the point itself is first discovered in the present
paper as well as that our study is the first numerical one which
comprehensively tackles three-flavour RSF effect with both the normal and
inverted mass hierarchy.

This paper is organized as following.
In \sref{sec:Formulation}, we give the formulation used in our
calculation, which includes all three-flavour neutrinos and
antineutrinos.
In \sref{sec:Models}, models of supernova neutrinos, presupernova
structure, and particle properties of the neutrino are illustrated.
In \sref{sec:Qualitative conversion schemes}, we give qualitative
discussions concerning neutrino conversions in supernova matter both for
the normal and inverted mass hierarchy, and show results of numerical
calculations in \sref{sec:Results of numerical calculations}.
Finally, in \sref{sec:Discussion}, a simple discussion how to obtain
information on the neutrino properties from the supernova neutrino
signal is presented, and dependence on presupernova models are
discussed.

\section{Formulation}
\label{sec:Formulation}

\subsection{Interaction with matter and magnetic fields}
\label{sub:Interaction with matter and magnetic fields}

The interaction of the magnetic moment of neutrinos and magnetic fields
is described by
\begin{equation}
\langle(\nu_{\rm i})_{\rm R}|H_{\mathrm{int}}|(\nu_{\rm j})_{\rm L}
 \rangle = \mu_{\rm ij} B_\perp ,
\label{eq:interaction Hamiltonian}
\end{equation}
where $\mu_{\rm ij}$ is the component of the neutrino magnetic moment
matrix, $B_\perp$ is the magnetic field transverse to the direction of
propagation, and $(\nu)_{\rm R}$ and $(\nu)_{\rm L}$ are the right- and
left-handed neutrinos, respectively.
If neutrinos are Dirac particles, right-handed neutrinos and
left-handed antineutrinos are undetectable (sterile neutrinos), since
they do not interact with matter.
On the other hand, if neutrinos are Majorana particles, $\nu_{\rm R}$
are identical to antiparticles of $\nu_{\rm L}$ and interact with
matter.
In this paper, we assume that neutrinos are Majorana particles.
The diagonal magnetic moments are forbidden for Majorana neutrinos,
and therefore only conversion between different flavours is
possible, e.g., $(\bar\nu_\rme)_{\rm R} \leftrightarrow
(\nu_{\mu,\tau})_{\rm L}$ or $(\nu_\rme)_L\leftrightarrow
(\bar\nu_{\mu,\tau})_{\rm R}$.

Coherent forward scattering with matter induces an effective
potential for neutrinos, which is calculated using weak interaction
theory.
The effective potential due to scattering with electrons is given by
\begin{equation}
V_{\pm\pm} = \pm \sqrt{2} G_{\rm F} \left( \pm \frac{1}{2} 
       + 2\sin^2 \theta_{\rm W} \right) n_\rme ,
\label{eq:electron potential}
\end{equation}
where $n_\rme $ is the electron number density, $G_{\rm F}$ is the
Fermi coupling constant, and $\theta_{\rm W}$ is the Weinberg angle.
The $\pm$ sign in front refers to $\nu~(+)$ and $\bar{\nu}~(-)$ and that
in the parentheses to $\nu_\rme ~(+)$ and $\nu_{\mu,\tau}~(-)$.
The difference between $e$ and $\mu,\tau$ neutrinos comes from the
existence of charged-current interaction.
The subscript $\pm\pm$ of $V$ refers to the first and the second $\pm$
sign.
The ordinary MSW effect between $\nu_\rme $ and $\nu_{\mu,\tau}$ is
caused by the potential difference $V_\rme -V_{\mu,\tau} =
V_{++}-V_{+-} = \sqrt{2}G_{\rm F} n_\rme $, while that between
$\bar\nu_\rme $ and $\bar\nu_{\mu,\tau}$ by $V_{\bar\rme }-V_
{\bar\mu,\bar\tau}=V_{-+}-V_{--}=-\sqrt{2}G_{\rm F}n_\rme $.
To include the RSF effect, which causes conversion between neutrinos
and antineutrinos, we should take into account the neutral-current
scattering by nucleons:
\begin{equation}
V = \sqrt{2} G_{\rm F} \left(\frac{1}{2} - 2 \sin^2 \theta_{\rm W}
 \right) n_{\rm p} - \sqrt{2} G_{\rm F} \frac{1}{2} n_{\rm n},
\label{eq:nucleon potential}
\end{equation}
where $n_{\rm p},n_{\rm n}$ are the proton and neutron number density,
respectively.
For neutrinos we add $+V$ to the potential and for antineutrinos $-V$.
Therefore, the RSF conversion between $\nu_\rme $ and
$\bar\nu_{\mu,\tau}$, which is important for the case considered in this
paper, obeys the potential difference
\begin{eqnarray}
\Delta V &\equiv& V_\rme -V_{\bar\mu,\bar\tau} \nonumber \\
 &=& ( V_{++} + V )  - ( V_{--} - V )  \nonumber \\
 &=& -\sqrt{2} G_{\rm F} \frac{\rho}{m_{\rm N}} ( 1 - 2 Y_\rme  ),
\label{eq:Delta V}
\end{eqnarray}
where $\rho$ is the density, $m_{\rm N}$ is the nucleon mass, and
$Y_\rme =n_\rme /(n_\rme +n_{\rm n})$ is the number of electrons
per baryon. (When we obtained equation \eref{eq:Delta V}, we assumed
charge neutrality $n_\rme =n_{\rm p}$.)

\subsection{The simplest case: $\nu_\rme \leftrightarrow
\bar\nu_\tau^\prime$ conversion}
\label{sub:simplest}

In this subsection, we give the simplest formulation between $\nu_\rme$
and $\bar\nu_\tau^\prime$ with the mass squared difference and mixing
angle between the first and third mass eigenstate $\Delta
m_{13}^2,\theta_{13}$, and consider the properties of the RSF
conversion.
Here, $\bar\nu_\tau^\prime$ represents mass eigenstate in matter which
can be obtained by some linear combination of $\bar\nu_\mu$ and
$\bar\nu_\tau$.
Although this treatment is obviously insufficient for a realistic
discussion, it is still useful for an intuitive understanding.
First, we qualitatively illustrate how the RSF conversions depend on the
neutrino mass hierarchy, and then more realistic three-flavour
formulation is given in the next subsection.
The time evolution of the mixed state of $\nu_\rme$ and $\bar\nu_\tau^
\prime$ is described by the Schr\"odinger equation
\begin{equation}
\rmi \frac{\rmd}{\rmd r} \left( 
	      \begin{array}{c}
	      \nu_\rme  \\
	      \bar\nu_\tau^\prime
	      \end{array}
	      \right)
= \left( 
   \begin{array}{cc}
    0 & -\mu_{\rme \tau^\prime} B_\perp \\
    -\mu_{\rme \tau^\prime} B_\perp & \Delta H
     \end{array}
 \right) \left( \begin{array}{c}
	\nu_\rme  \\
	\bar\nu_\tau^\prime
	\end{array} \right),
\label{eq:two-flavour}
\end{equation}
where $r$ is the radius from the center of the star, $\mu_{\rme \tau^
\prime}$ is the transition magnetic moment, and $\Delta H$ is defined by
\begin{equation}
\Delta H \equiv \frac{\Delta m^2_{13}} 
{2E_\nu} \cos 2\theta_{13} - \Delta V.
\label{eq:Delta H}
\end{equation}
The resonance occurs when $\Delta H=0$; it does not occur if $\Delta
m_{13}^2>0$ (normal hierarchy), because $Y_\rme <0.5$ (therefore,
$\Delta V<0$) is satisfied in stellar envelope.
Instead, the conversion between $\bar\nu_\rme$ and $\nu_\tau$ is
affected by the resonance since the potential difference between them is
$V_{\bar\rme}-V_{\mu,\tau}=-\Delta V(>0)$.
On the other hand, in the case of the inverted hierarchy, the RSF
conversion takes place between $\nu_\rme$ and $\bar\nu_\tau^\prime$, but
not between $\bar\nu_\rme$ and $\nu_\tau^\prime$.
This situation is very similar to the case of pure MSW effect; higher
MSW resonance occurs in antineutrino (neutrino) sector if the mass
hierarchy is inverted (normal) \cite{Dighe00,Takahashi03a}.

When the resonance is adiabatic (or when the density is slowly changing
at that point and the magnetic field is strong enough), significant
conversion occurs. 
The adiabaticity of the RSF conversion is given by
\begin{equation}
\gamma^{\mathrm{RSF}}_{\rme \bar\tau^\prime} =
 \left.\frac{(2\mu_{\rme \tau^\prime}B_\perp)^2}
 {|\rmd\Delta V / \rmd r|}\right|_{\mathrm{res}},
\label{eq:adiabaticity}
\end{equation}
where the subscript ``res'' means that the value is evaluated at the
resonance point.
Therefore, if the magnetic field is sufficiently strong at the resonance
point, the $\nu_\rme \leftrightarrow \bar\nu_\tau^\prime$ conversion
occurs completely.

\subsection{Three-flavour formulation}
\label{sub:Three-flavour formulation}

Here, we present the three-flavour (six-component) formulation of
neutrino mixing, on which our discussions depend:
\begin{equation}
\rmi \frac{\rmd}{\rmd r} \left(
	      \begin{array}{c} 
	       \nu \\
	       \bar{\nu}
	      \end{array}
	      \right)
= \left( \begin{array}{cc}
 H_0 & B_\perp M \\ -B_\perp M & \bar{H_0} \end{array} \right)
\left( \begin{array}{c} \nu \\ \bar{\nu} \end{array} \right),
\label{eq:three-flavour}
\end{equation}
where 
\begin{equation}
\fl\nu = \left(
	 \begin{array}{c}
	  \nu_\rme  \\
	  \nu_\mu \\
	  \nu_\tau
	 \end{array}
       \right), 
~~\bar{\nu} = \left(
	 \begin{array}{c}
	  \bar{\nu}_\rme \\
	  \bar{\nu}_\mu \\
	  \bar{\nu}_\tau
	 \end{array}
       \right), 
\end{equation}
 \begin{eqnarray}
  \fl H_0 = \frac{1}{2E_\nu} U \left(
			  \begin{array}{ccc}
			  0 & 0 & 0 \\
			  0 & \Delta m^2_{12} & 0 \\
			  0 & 0 & \Delta m^2_{13}
			  \end{array}
			  \right) U^\dagger
  + \left(
   \begin{array}{ccc}
   V_{++}+V & 0 & 0 \\
   0 & V_{+-}+V & 0 \\
   0 & 0 & V_{+-}+V 
   \end{array}
   \right), \label{equationa} 
\\
  \fl \bar{H_0} = \frac{1}{2E_\nu} U \left(
			  \begin{array}{ccc}
			  0 & 0 & 0 \\
			  0 & \Delta m^2_{12} & 0 \\
			  0 & 0 & \Delta m^2_{13}
			  \end{array}
			  \right) U^\dagger
  + \left(
   \begin{array}{ccc}
   V_{-+}-V & 0 & 0 \\
   0 & V_{--}-V & 0 \\
   0 & 0 & V_{--}-V 
   \end{array}
   \right), \label{equationb}
\\
  \fl U = \left(
       \begin{array}{ccc}
	U_{\rme 1} & U_{\rme 2} & U_{\rme 3} \\
        U_{\mu 1} & U_{\mu 2} & U_{\mu 3} \\
        U_{\tau 1} & U_{\tau 2} & U_{\tau 3}
       \end{array}
     \right)
  = \left(
   \begin{array}{ccc}
   c_{12}c_{13} 
    & s_{12}c_{13} 
   & s_{13} \\
   -s_{12}c_{23}
    -c_{12}s_{23}s_{13}
  & c_{12}c_{23}
  -s_{12}s_{23}s_{13} 
  & s_{23}c_{13} \\
    s_{12}s_{23}
     -c_{12}c_{23}s_{13}
     & -c_{12}s_{23}
     -s_{12}c_{23}s_{13}
     & c_{23}c_{13}
     \end{array}
   \right), \label{eq:U}
 \end{eqnarray}
\begin{equation}
\fl M = \left(
   \begin{array}{ccc}
   0 & \mu_{\rme \mu} & \mu_{\rme \tau} \\
   -\mu_{\rme \mu} & 0 & \mu_{\mu\tau} \\
   -\mu_{\rme \tau} & -\mu_{\mu\tau} & 0
   \end{array}
   \right),
\label{eq:magnetic moment}
\end{equation}
and $c_{\rm ij}=\cos\theta_{\rm ij},~s_{\rm ij}=\sin\theta_{\rm ij}$.
(We assume the CP phase $\delta=0$ in equation \eref{eq:U} for
simplicity.)

The resonant flavour conversion basically occurs when the two diagonal
elements in the matrix in equation \eref{eq:three-flavour} have the same
value.
There are four relevant resonance points, but they depends on the
neutrino mass hierarchy as discussed in the previous subsection.
In the case of the normal mass hierarchy, they are for $\nu_\rme 
\leftrightarrow \nu_\mu^\prime$ (MSW-L), $\nu_\rme  \leftrightarrow
\nu_\tau^\prime$ (MSW-H), $\bar\nu_\rme  \leftrightarrow
\nu_\mu^\prime$ (RSF-L), and $\bar\nu_\rme \leftrightarrow
\nu_\tau^\prime$ (RSF-H).
Here, the quantities such as $\nu_{\mu,\tau}^\prime (\bar\nu_{\mu,\tau}
^\prime)$ represent mass eigenstates in matter which can be obtained by
linear combination of $\nu_\mu (\bar\nu_\mu)$ and $\nu_\tau
(\bar\nu_\tau)$.
This is because the $\nu_\rme (\bar\nu_\rme)$ state coincides with mass
eigenstate in matter owing to large matter potential, and the other
mass eigenstates are obtained by the rotation of
$\nu_\mu(\bar\nu_\mu)$ and $\nu_{\tau}(\bar\nu_\tau)$ basis.
Suffixes `-L' and `-H' attached to `MSW' and `RSF' indicate whether the
density at the resonance points are lower or higher.
On the other hand, in the case of the inverted hierarchy, RSF-H occurs
for $\nu_\rme \leftrightarrow\bar\nu_\tau^\prime$ as well as MSW-H for
$\bar\nu_\rme \leftrightarrow\bar\nu_\tau^\prime$, whereas the other
two resonances occurs for the same conversions.
This situation is summarized in \tref{table:resonance}.

\begin{table}
\caption{Flavour conversions that are important for each resonance, in
 both cases of the normal and inverted mass
 hierarchy. \label{table:resonance}}
\begin{indented}
\item[]\begin{tabular}{@{}lll}
\br
Resonance & Normal hierarchy & Inverted hierarchy\\
\mr
RSF-H & $\bar\nu_\rme \leftrightarrow\nu_\tau^\prime$
& $\nu_\rme \leftrightarrow\bar\nu_\tau^\prime$\\
RSF-L & $\bar\nu_\rme \leftrightarrow\nu_\mu^\prime$
& $\bar\nu_\rme \leftrightarrow\nu_\mu^\prime$\\
MSW-H & $\nu_\rme \leftrightarrow\nu_\tau^\prime$
& $\bar\nu_\rme \leftrightarrow\bar\nu_\tau^\prime$\\
MSW-L & $\nu_\rme \leftrightarrow\nu_\mu^\prime$
& $\nu_\rme \leftrightarrow\nu_\mu^\prime$\\
\br
\end{tabular}
\end{indented}
\end{table}

\section{Models}
\label{sec:Models}

\subsection{Original neutrino emission}
\label{sub:Original neutrino emission}

\begin{figure}[htbp]
\begin{center}
\includegraphics[width=10cm]{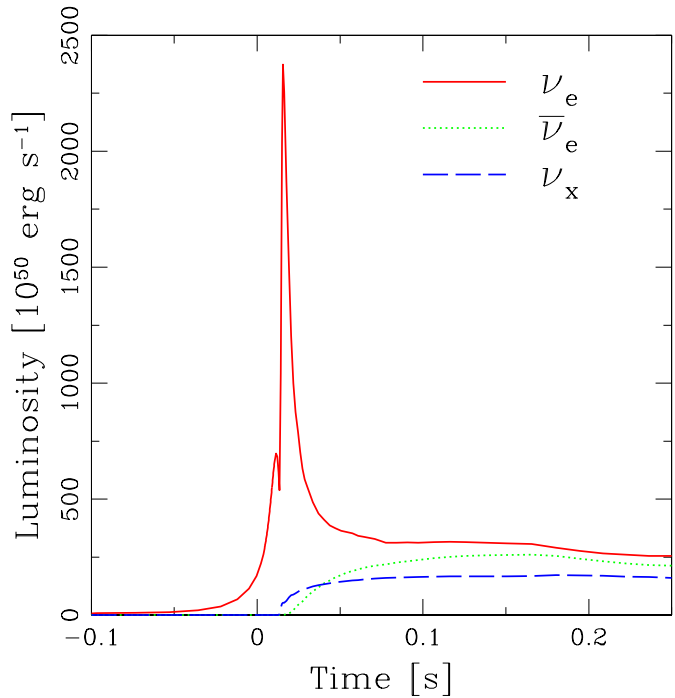}
\caption{The original luminosity of the emitted neutrinos as a function
 of time, calculated by Thompson \etal \cite{Thompson03}. The progenitor
 mass is $15M_\odot$. \label{fig:luminosity_curve_org}}
\end{center}
\end{figure}

\begin{figure}[htbp]
\begin{center}
\includegraphics[width=10cm]{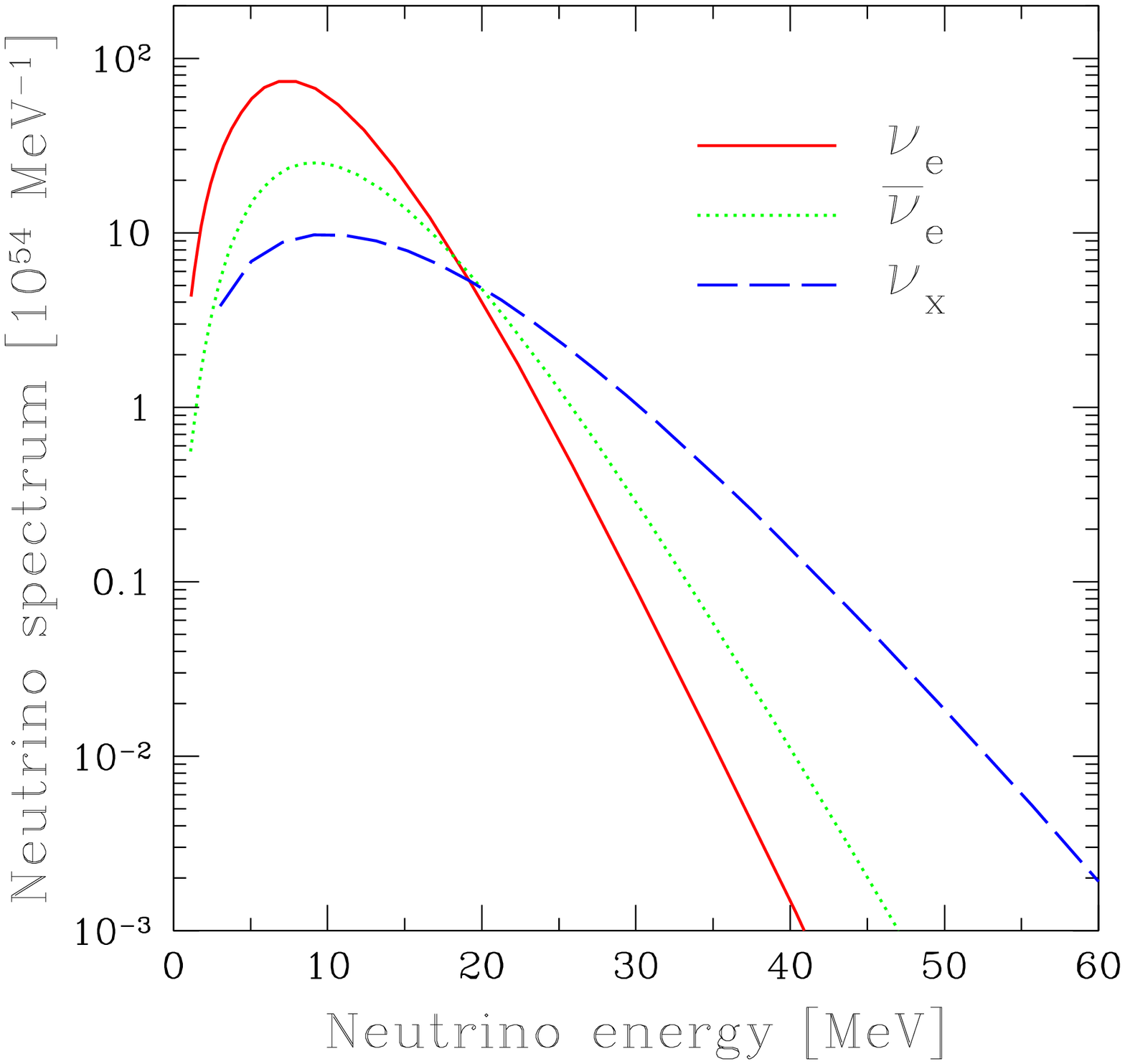}
\caption{Original neutrino spectrum integrated to 0.25 s after core
 bounce, calculated by Thompson \etal \cite{Thompson03}. The progenitor
 mass is $15M_\odot$. \label{fig:spectrum_org}}
\end{center}
\end{figure}

We adopt, as original neutrino spectrum as well as luminosity curve, the
result of numerical simulation by Thompson \etal \cite{Thompson03}; we
use the model calculated for the $15M_\odot$ progenitor star.
Their calculation has particularly focused on shock breakout and
followed the dynamical evolution of the cores through collapse until the
first 250 ms after bounce.
They have incorporated all the relevant neutrino processes such as
neutrino--nucleon scatterings with nucleon recoil as well as nucleon
bremsstrahlung; these reactions have recently been recognized to give
non-negligible contribution to the spectral formation.
In figures \ref{fig:luminosity_curve_org} and \ref{fig:spectrum_org}, we
show the original luminosity curve and number spectrum of neutrinos,
respectively.
In these figures, $\nu_{\rm x}$ represents non-electron neutrinos and
antineutrinos.

Neutrino luminosity curve is quite characteristic among different
flavours.
In particular, there is a very sharp peak of $\nu_\rme$ called
`neutronization burst', whose duration is typically $\sim 10$ ms and
peak luminosity is $\sim 10^{53}$ erg/s.
This strong peak is illustrated as follows.
As a supernova shock moves outward, it dissociates nuclei into free
nucleons, which triggers deleptonization process $\rme^-{\rm p}\to\nu
_\rme {\rm n}$;
these $\nu_\rme$ build up a sea because they are trapped and advected
with matter.
When the shock crosses the $\nu_\rme$ neutrinosphere, within which
the created $\nu_\rme$ are trapped, they are abruptly emitted.
For the other flavours $\bar\nu_\rme$ and $\nu_{\rm x}$, there is no
such a sudden burst; both luminosities glow rather gradually and they
are similar to each other.
Usually, for both the pure MSW effect and the RSF effect with normal
hierarchy, the most easily detected flavour $\bar\nu_\rme$ are
transformed from original $\bar\nu_\rme$ and $\nu_{\rm x}$
\cite{Ando03b,Ando03d}.
Thus, luminosity curve does not provide any useful information on the
flavour conversion mechanism.
On the other hand, for the RSF effect with inverted mass hierarchy, the
conversion $\nu_\rme\to\bar\nu_\tau^\prime\to\bar\nu_\rme$ is considered
to occur via RSF-H and MSW-H (see \tref{table:resonance}); we expect
that this case can be distinguished from the luminosity curve.

The other characteristic that provides information on flavour conversion
mechanism is hierarchy of the average energy $\langle
E_{\nu_\rme}\rangle <\langle E_{\bar\nu_\rme }\rangle <\langle
E_{\nu_{\rm x}}\rangle$ as clearly seen from \fref{fig:spectrum_org};
flavour conversions also cause the spectral exchange.
This energy hierarchy is explained as follows.
Since $\nu_{\rm x}$ interact with matter only through the
neutral-current interactions in supernovae, they are weakly coupled with
matter compared to $\nu_\rme$ and $\bar\nu_\rme$.
Thus the neutrinosphere of $\nu_{\rm x}$ locates at deeper in the core
than that of $\nu_\rme$ and $\bar\nu_\rme$, which leads to higher
temperatures for $\nu_{\rm x}$.
The difference between $\nu_\rme$ and $\bar\nu_\rme$ comes from the fact
that the core is neutron-rich and $\nu_\rme$ couple with matter more
strongly, through $\nu_\rme {\rm n}\to \rme^-{\rm p}$ reaction.

\subsection{Presupernova profiles}
\label{sub:Presupernova profiles}

\begin{figure}[htbp]
\begin{center}
\includegraphics[width=10cm]{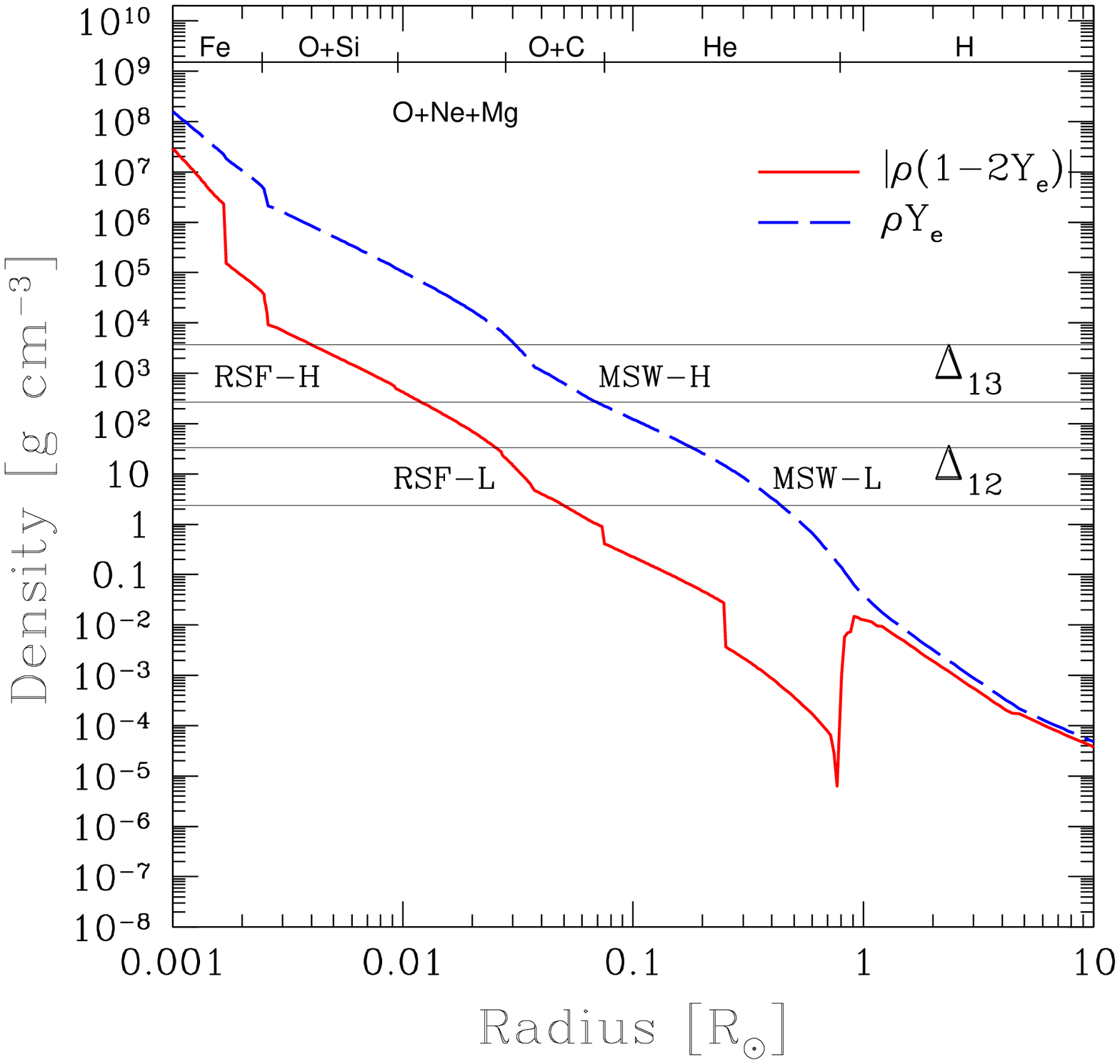}
\caption{Presupernova profiles (W95S) used in our calculations, which is
 calculated by Woosley and Weaver \cite{Woosley95}. The density and
 $Y_\rme $ combination that is responsible for the RSF conversions
 [$\rho (1-2Y_\rme )$], and that for the MSW conversions ($\rho Y_\rme$).
 Two horizontal bands represent $\Delta_{12}$ and $\Delta_{13}$ (these
 definitions are given in the text); at the intersections between them
 and the profile curves, the RSF and MSW conversions
 occur. \label{fig:profiles}}
\end{center}
\end{figure}

We use the precollapse model of massive stars of Woosley and Weaver
\cite{Woosley95}.
The model is $15M_\odot$ progenitor star with solar metallicity and it
is labeled as W95S.
The density and $Y_\rme$ profiles are quite important for the flavour
conversions because they determine the resonance regions as well as
whether it is adiabatic or not.
We show in \fref{fig:profiles} the $|\rho (1-2Y_\rme)|$ (responsible
for RSF) and $\rho Y_\rme$ profiles (responsible for MSW) of the W95S
model.
We also show $\Delta_{12}\equiv m_{\rm N}\Delta m_{12}^2\cos
2\theta_{12}/2\sqrt 2G_{\rm F}E_\nu$ and $\Delta_{13}\equiv m_{\rm
N}|\Delta m_{13}^2|\cos 2\theta_{13}/2\sqrt 2G_{\rm F}E_\nu$ as two
horizontal bands (the bandwidth comes from the energy range 5--70 MeV).
At intersections between $\Delta_{12},\Delta_{13}$ and $|\rho
(1-2Y_\rme)|, \rho Y_\rme $, the RSF and MSW conversions take place.

\begin{figure}[htbp]
\begin{center}
\includegraphics[width=10cm]{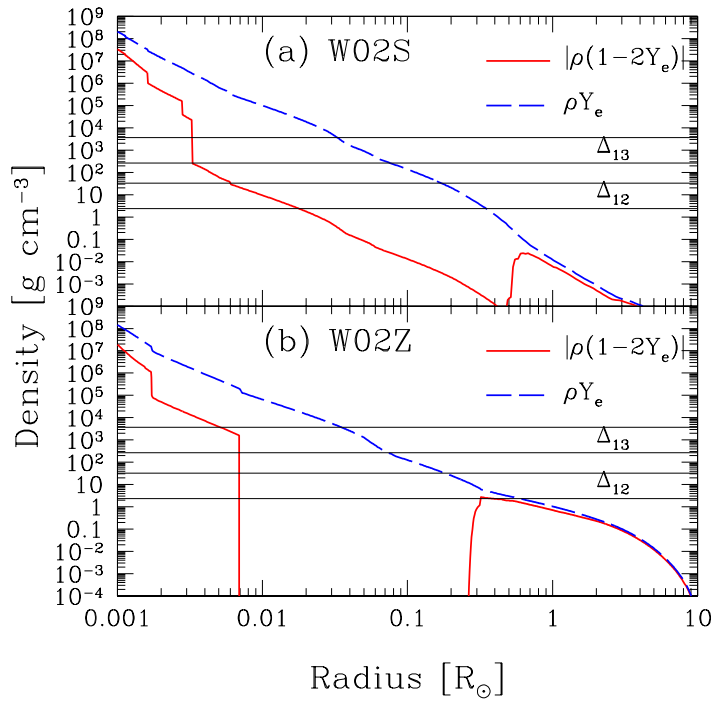}
\caption{The same as \fref{fig:profiles} but for (a) W02S and (b) W02Z
 models. \label{fig:profiles_W02}}
\end{center}
\end{figure}

In the previous publication, we have investigated the dependence on
presupernova models \cite{Ando03d}; in the paper, we have also used
recent $15M_\odot$ presupernova models by Woosley \etal \cite{Woosley02}
with both solar and zero metallicity (W02S and W02Z, where `S' and
`Z' denotes solar and zero metallicity, respectively).
This is because the RSF conversion is very sensitive to the deviation of
$Y_\rme$ from 0.5 (see equation \eref{eq:Delta V}), which strongly
depends on the metallicities as well as on the weak interaction rates
adopted in the simulation of stellar evolution.
In fact, for the W02 models, in which the authors adopted more updated
weak rates, the $|\rho (1-2Y_\rme)|$ profiles suddenly drop at RSF
region, yielding rather nonadiabatic conversions; these properties are
shown in \fref{fig:profiles_W02}.
Nevertheless, we use the W95S model as our reference model because of
the reasons listed below.
First, since the conversion probabilities in RSF-H is highly adiabatic
when the magnetic field is sufficiently strong (\sref{sec:Results of
numerical calculations}), it is easier for us to investigate the
dependence on mass hierarchy.
Second, although the W02 models were calculated using a recent shell
model and resulted in substantial revisions to the older data sets in
the W95S model, there are still many uncertainties concerning nuclear
physics; thus, these models cannot be considered to be decisive one.
Finally, as discussed in the previous paper \cite{Ando03d}, because the
lifetime of massive stars which end their life by gravitational collapse
is much shorter than that of the Sun, the progenitors of the galactic
supernovae are considered to be younger than the Sun, and therefore,
more metal rich.
Because the deviation of $Y_\rme$ from 0.5 is determined by rarely
existent nuclei, the suppression of $(1-2Y_\rme)$ will be weaker and the
RSF conversion will incline to be more adiabatic, even if W02S model is
the correct one of the progenitor with solar metallicity.

Although we use static progenitor models in calculating the flavour
transition, in fact the density profile changes drastically during a
neutrino burst ($\sim 10$ s) owing to shock wave propagation.
However, our particular interest is within first 0.25 s after core
bounce because the calculation of original neutrino emission
\cite{Thompson03} ends at that time (\ref{sub:Original neutrino
emission}).
In the previous paper, we have already shown that until 0.25 s, using
the static presupernova and magnetic field models is considered to be a
good approximation \cite{Ando03b}; this is based on the numerical
calculation of the Lawrence Livermore group, which is the only group
succeeding the shock propagation into the outer envelope.

\subsection{Magnetic fields}
\label{sub:Magnetic fields}

We assume that the global structure of the magnetic field is a dipole
moment and the field strength is normalized at the surface of the iron
core with the values $10^{10}$ G (nearly complete RSF conversion) as
well as 0 G (pure MSW conversion).
The reason for this normalization is as follows.
The magnetic fields should be normalized by fields that are static
and exist before the core collapse, because those of a nascent
neutron star can hardly affect the far outer region, where the RSF
conversions take place, within the short time scale of a neutrino burst.
As discussed in the previous subsection, since the shock wave does not
affect the resonance region at $\lesssim 0.25$ s after bounce, it is
also expected that the magnetic field structure and strength at the
resonance points are not seriously changed at that time.
The strength of such magnetic fields above the surface of the iron core
may be inferred from observations of the surface of white dwarfs, since
both are sustained against gravitational collapse by the degenerate
pressure of electrons.
Observations of the magnetic fields in white dwarfs show that the
strength spreads in a wide range of $10^7$--$10^9$ G
\cite{Chanmugam92}.
Considering the possibility of the decay of magnetic fields in white
dwarfs, it is not unnatural to consider magnetic fields up to
$10^{10}$ G at the surface of the iron core.
Then, in equation \eref{eq:three-flavour}, $B_\perp = B_0 (r_0 / r)^3
\sin\Theta$, where $B_0$ is the strength of the magnetic field at the
equator on the iron core surface, $r_0$ the radius of the iron core, and
$\Theta$ the angle between the pole of the magnetic dipole and the
direction of neutrino propagation. 
Hereafter, we assume $\sin \Theta = 1$.

\subsection{Neutrino parameters}
\label{sub:Neutrino parameters}

We adopt the realistic neutrino mixing parameters inferred from the
recent experimental results: for the atmospheric neutrino parameters,
$|\Delta m_{13}^2|=2.8\times 10^{-3}~{\rm eV}^2,\sin^22\theta_{23}=1.0$,
and for the solar neutrino parameters, $\Delta m_{12}^2=5.0\times
10^{-5}~{\rm eV}^2,\tan^2\theta_{12}=0.42$.

As for still uncertain neutrino properties, we must set some
assumptions, e.g., whether the mass hierarchy is normal or inverted, as
well as $\theta_{13}$ is large, enough for the MSW-H conversion to be
adiabatic, or not.
There is also uncertainty concerning the neutrino magnetic moment tensor
$\mu_{\rm ij}$; however, since the only relevant parameter is $\mu_{\rm
ij}B$, this uncertainty is already included in that of the magnetic
field strength ($\mu_{\rm ij}=10^{-12}\mu_B$ is assumed).
Thus, there are 8 parameter sets due to:
whether the magnetic field is zero (label by `MSW') or sufficiently
strong $B_0=10^{10}$ G (`RSF');
the mass hierarchy is normal (`NOR') or inverted (`INV');
and $\sin^22\theta_{13}=10^{-6}$ (`S') or $\sin^22\theta_{13}=0.04$
(`L').
We label one model by connecting these labels using hyphen, e.g.,
MSW-NOR-S.
It should be particularly noticed that although we label the model with
strong magnetic field simply by RSF-, it does not mean that the pure RSF
effect occurs; every model labeled by RSF- is subject to both the MSW
and RSF conversions.
We summarize these models in \tref{table:models}.

\begin{table}
\caption{Models considered in this paper, concerning the neutrino
 properties. \label{table:models}}
\begin{indented}
\item[]\begin{tabular}{@{}llll}
\br
Model & $B_0$ [G] & Mass hierarchy & $\sin^22\theta_{13}$\\
\mr
MSW-NOR-S & 0 & Normal & $10^{-6}$\\
MSW-NOR-L & 0 & Normal & 0.04\\
MSW-INV-S & 0 & Inverted & $10^{-6}$\\
MSW-INV-L & 0 & Inverted & 0.04\\
RSF-NOR-S & $10^{10}$ & Normal & $10^{-6}$\\
RSF-NOR-L & $10^{10}$ & Normal & 0.04\\
RSF-INV-S & $10^{10}$ & Inverted & $10^{-6}$\\
RSF-INV-L & $10^{10}$ & Inverted & 0.04\\
\br
\end{tabular}
\end{indented}
\end{table}

\section{Qualitative conversion schemes}
\label{sec:Qualitative conversion schemes}

\begin{figure}[htbp]
\begin{center}
\includegraphics[width=15cm]{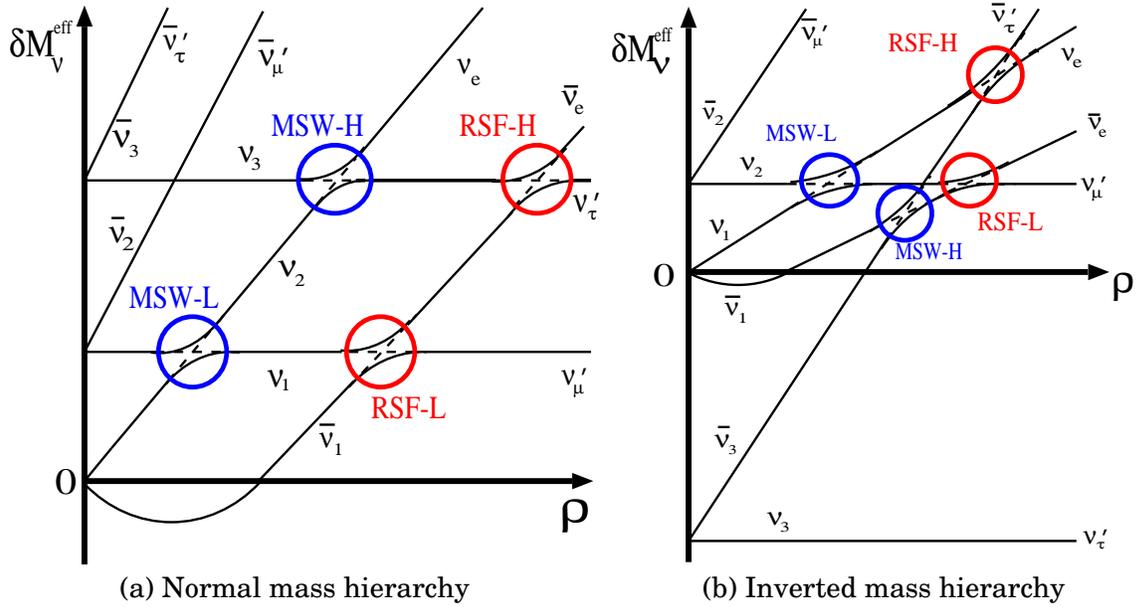}
\caption{Schematic illustration of level crossings for (a) normal and
 (b) inverted mass hierarchy. In this figure, adiabatic conversion means
 that the neutrinos trace the solid curve at each resonance point, while
 nonadiabatic conversion is shown by the dashed
 line.\label{fig:crossing}}
\end{center}
\end{figure}

We qualitatively illustrate the conversion scheme for each model.
Since $\bar\nu_\rme$ is the most easily detected flavour, we focus on the
conversion channel that gives $\bar\nu_\rme$ appearance at the detector.
\Fref{fig:crossing} schematically shows crossings among different mass
eigenstates in matter, which is helpful for the reader to understand the
qualitative discussions in this section.

\subsection{Normal mass hierarchy}
\label{sub:Normal mass hierarchy}

First, the case of the normal mass hierarchy is addressed
(\fref{fig:crossing}(a)); this case is minutely investigated in
\cite{Ando03b,Ando03d}, but we again repeat their discussion.
When $B_0=0$, the produced $\bar\nu_\rme$ propagate the supernova envelope
without experiencing the MSW resonance, and reach the stellar surface as
the lightest mass eigenstate $\bar\nu_1$.
The other mass eigenstates $\bar\nu_2,\bar\nu_3$ are originated from
$\bar\nu_{\mu,\tau}$ whose flux is considered to be the same at the
leading order.
Thus, the $\bar\nu_\rme$ flux at the detector is described by
\begin{eqnarray}
 F_{\bar\nu_\rme}&=&|U_{\rme 1}|^2F_{\bar\nu_1}+|U_{\rme 2}|^2
  F_{\bar\nu_2}+|U_{\rme 3}|^2F_{\bar\nu_3}\nonumber\\
 &=&|U_{\rme 1}|^2F_{\bar\nu_\rme}^0+(1-|U_{\rme 1}|^2)
  F_{\nu_{\rm x}}^0,
  \label{eq:flux for group A}
\end{eqnarray}
where $F^0$ and $F$ are the flux at production and detection,
respectively.
Since the parameter $\theta_{13}$ is essential for the MSW-H conversion,
the results are not sensitive to the value of $\theta_{13}$.

On the other hand, when $B_0=10^{10}$ G, RSF-H becomes almost completely
adiabatic, whereas RSF-L is highly nonadiabatic.
At the RSF-H point, the original $\nu_\tau^\prime$ are converted into
$\bar\nu_\rme$, which propagate as the mass eigenstate owing to the
large matter potential.
These $\bar\nu_\rme$, then, cross the nonadiabatic RSF-L region (which
gives no essential effects) and escape from the star as $\bar\nu_1$.
The other mass eigenstates are also originated from $\nu_{\rm x}$, thus
yielding
\begin{equation}
 F_{\bar\nu_\rme}=F_{\nu_{\rm x}}^0.
  \label{eq:flux for group B}
\end{equation}
The value of $\theta_{13}$ does not matter also in this case.
In consequence, the models MSW-NOR-S,L are characterized by equation
\eref{eq:flux for group A}, whereas, RSF-NOR-S,L are characterized by
equation \eref{eq:flux for group B}.

\subsection{Inverted mass hierarchy}
\label{sub:Inverted mass hierarchy}

The situation changes dramatically in the case of the inverted mass
hierarchy (\fref{fig:crossing}(b)).
The pure MSW effect in this case has been studied in detail in the
literatures \cite{Dighe00,Takahashi03a}; the value of $\theta_{13}$ is
critical for completeness of the conversions.
The $\bar\nu_\rme$, which are produced as the lightest mass eigenstate
in three antineutrino states, then cross the MSW-H resonance region.
If this resonance is nonadiabatic (MSW-INV-S), the flavour conversion
does not take place at the MSW-H point, i.e.,
$\bar\nu_\rme\to\bar\nu_1$.
This case yields the $\bar\nu_\rme$ flux characterized by equation
\eref{eq:flux for group A}, which has already appeared.
On the other hand for the MSW-INV-L model, MSW-H is completely
adiabatic, and therefore the conversions $\bar\nu_\rme\to\bar\nu_\tau^
\prime\to\bar\nu_3$ as well as $\nu_{\rm x}\to\bar\nu_{1,2}$ occur,
which results in
\begin{equation}
 F_{\bar\nu_\rme}=(1-|U_{\rme 3}|^2)F_{\nu_{\rm x}}^0
  +|U_{\rme 3}|^2F_{\bar\nu_\rme}^0.
\end{equation}
Since the value of $\theta_{13}$ (or $|U_{\rme 3}|^2$) is strongly
constrained to be very small by the reactor experiment
\cite{Apollonio99}, this expression essentially the same as the previous
one \eref{eq:flux for group B}.

When $B_0=10^{10}$ G (RSF-INV-S,L), the $\nu_\rme\leftrightarrow\bar\nu_
\tau^\prime$ conversion takes place at the RSF-H region.
This $\bar\nu_\tau^\prime$ further enter the MSW-H region.
If this MSW resonance is nonadiabatic (-S model), these $\bar\nu_\tau^
\prime$ becomes $\bar\nu_3$ at the stellar surface.
The other relevant conversions are the same as the pure MSW model
$\bar\nu_\rme\to\bar\nu_1,\bar\nu_\mu^\prime\to\bar\nu_2$.
Thus, the $\bar\nu_\rme$ flux is given by
\begin{equation}
 F_{\bar\nu_\rme}=|U_{\rme 1}|^2F_{\bar\nu_\rme}^0+|U_{\rme 2}|^2
  F_{\nu_{\rm x}}^0 +|U_{\rme 3}|^2F_{\nu_\rme}^0,
\end{equation}
which is almost the same as equation \eref{eq:flux for group A}.
On the other hand, for -L model, the mass eigenstates $\bar\nu_\tau^
\prime$ converted from $\nu_\rme$ reach the supernova surface as
$\bar\nu_1$, because the adiabatic MSW-H conversion $\bar\nu_\tau^
\prime\leftrightarrow\bar\nu_\rme$ takes place.
With the remaining channels $\bar\nu_\mu^\prime\to\bar\nu_2,\bar\nu_\rme
\to\bar\nu_3$, the observed flux is
\begin{equation}
 F_{\bar\nu_\rme}=|U_{\rme 1}|^2F_{\nu_\rme}^0+|U_{\rme 2}|^2
  F_{\nu_{\rm x}}+|U_{\rme 3}|^2F_{\bar\nu_\rme}^0.
  \label{eq:flux for group C}
\end{equation}
Consequently, the models MSW-INV-S and RSF-INV-S are characterized by
equation \eref{eq:flux for group A}, MSW-INV-L is by \eref{eq:flux for
group B} and RSF-INV-L is by \eref{eq:flux for group C}.

In \tref{table:conversions}, we summarize the relevant conversions
stated above for each model.
Although there are eight models and this rather large number is expected
to complicate the discussion, each model is consequently categorized
into one of three groups A, B or C, as indicated in
\tref{table:conversions}.
Fortunately, this greatly reduces the complexity, but within one group,
we cannot specify each model by the SK observation.

\begin{table}
\caption{Conversion scheme for each model. In the second column, the
 most relevant conversion channel, which eventually results in
 $\bar\nu_1$, is given. The flux equation which is relevant for each
 model is summarized in the third column, and then the models are
 categorized into groups labeled by A, B, and C, as shown in the fourth
 column. \label{table:conversions}}
\begin{indented}
\item[]\begin{tabular}{@{}llll}
\br
Model & Relevant conversion & Flux at the Earth & Group\\
\mr
MSW-NOR-S & $\bar\nu_\rme\to\bar\nu_1$ & \Eref{eq:flux for group A} & A\\
MSW-NOR-L & $\bar\nu_\rme\to\bar\nu_1$ & \Eref{eq:flux for group A} & A\\
MSW-INV-S & $\bar\nu_\rme\to\bar\nu_1$ & \Eref{eq:flux for group A} & A\\
MSW-INV-L & $\bar\nu_\tau^\prime\to\bar\nu_\rme \to\bar\nu_1$ &
	\Eref{eq:flux for group B} & B\\
RSF-NOR-S & $\nu_\tau^\prime\to\bar\nu_\rme\to\bar\nu_1$ & \Eref{eq:flux
	for group B} & B\\
RSF-NOR-L & $\nu_\tau^\prime\to\bar\nu_\rme\to\bar\nu_1$ & \Eref{eq:flux
	for group B} & B\\
RSF-INV-S & $\bar\nu_\rme\to\bar\nu_1$ & \Eref{eq:flux for group A} & A\\
RSF-INV-L & $\nu_\rme\to\bar\nu_\tau^\prime\to\bar\nu_\rme\to\bar\nu_1$ &
	\Eref{eq:flux for group C} & C\\
\br
\end{tabular}
\end{indented}
\end{table}

\section{Results of numerical calculations}
\label{sec:Results of numerical calculations}

\subsection{Conversion probability}
\label{sec:Conversion probability}

\begin{figure}[htbp]
\begin{center}
\includegraphics[width=10cm]{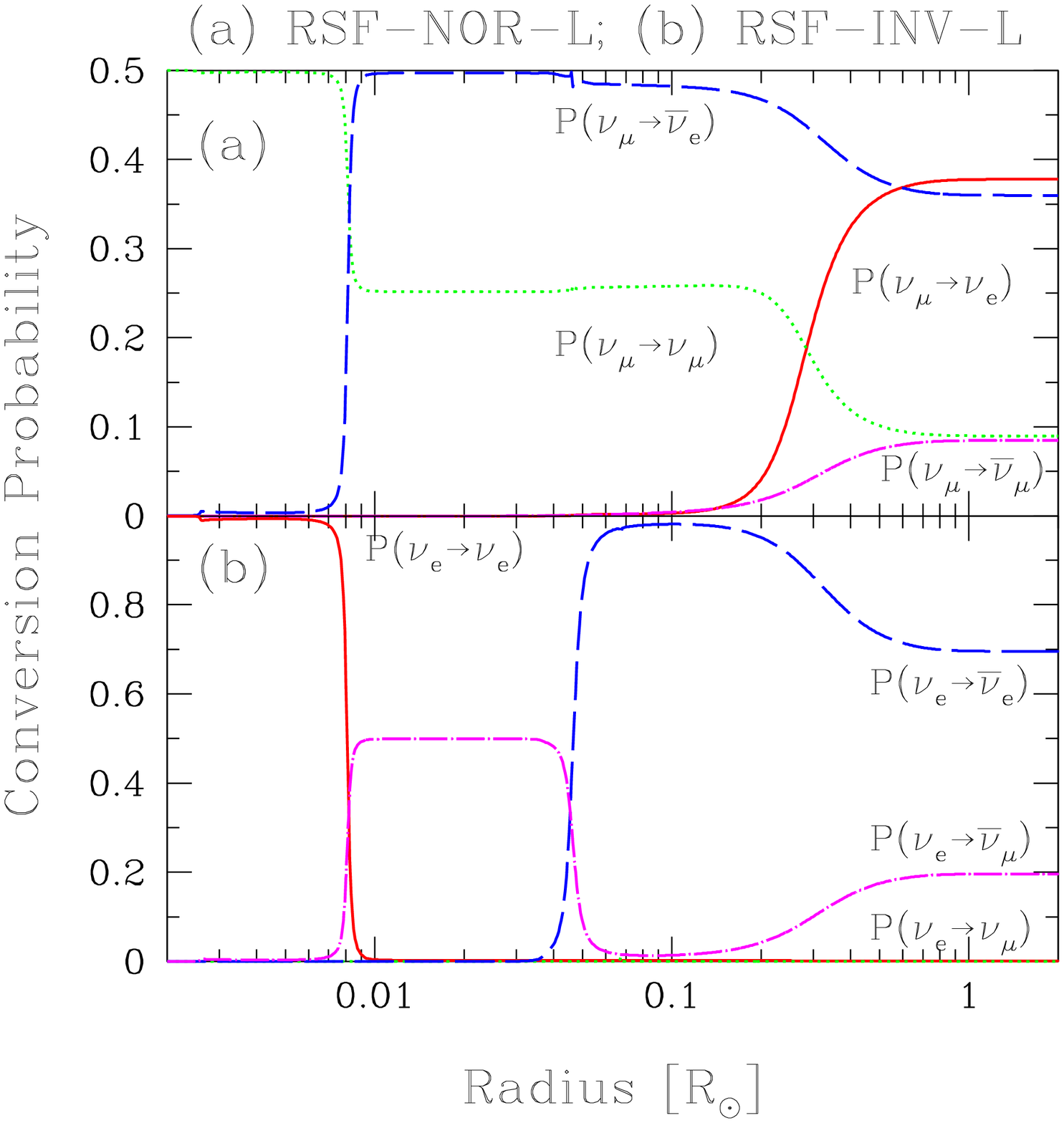}
\caption{Conversion probabilities as a function of radius for (a) the
 RSF-NOR-L and (b) RSF-INV-L models. The injected neutrino energy is
 taken to be 25 MeV. \label{fig:prob_L}}
\end{center}
\end{figure}

We calculated equation \eref{eq:three-flavour} numerically with the
models given in \sref{sec:Models}, and obtained the conversion
probabilities for each flavour.
\Fref{fig:prob_L}(a) shows the conversion probabilities of original
$\nu_\mu$ as a function of radius for the RSF-NOR-L model; the neutrino
energy is 25 MeV.
Because the $\nu_\mu$ state is not mass eigenstate and maximally mixing
with $\nu_\tau$, the probability $P(\nu_\mu\to\nu_\mu)$ should
oscillate abruptly.
In our calculation, however, we have averaged out this local behaviour,
since we are only interested in the global change of the probabilities
which is concerned with resonance.
Therefore, the $\nu_\mu$ survival probability starts from 0.5 rather
than 1, because of the maximal mixing with $\nu_\tau$.
A significant amount of $\nu_\mu$ change into $\bar\nu_\rme$, owing
mainly to the RSF-H conversion which occurs around $\sim 0.01R_\odot$,
and the converted $\bar\nu_\rme$ propagates as the mass eigenstate due
to matter effect, not being disturbed by further resonances, to $\sim
0.1R_\odot$.
Around this radius, $\bar\nu_\rme$ start to mix with other flavour
antineutrinos, reducing the probability $P(\nu_\mu\to\bar\nu_\rme)$.
In \fref{fig:prob_L}(b), we show the conversion probabilities of
original $\nu_\rme$ for the RSF-INV-L model.
As already discussed in \ref{sub:Inverted mass hierarchy}, the original
$\nu_\rme$ are transformed into $\bar\nu_{\mu,\tau}$ at RSF-H, and they
further change to the favoured flavour $\bar\nu_\rme$ in the MSW-H
resonance point.

\begin{figure}[htbp]
\begin{center}
\includegraphics[width=10cm]{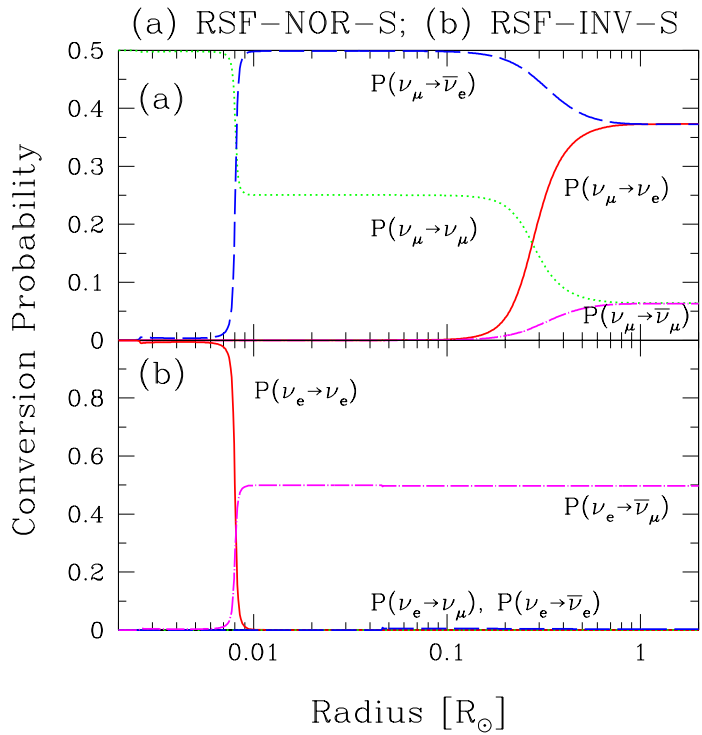}
\caption{The same as \fref{fig:prob_S}, but for (a) the RSF-NOR-S and
 (b) RSF-INV-S models. \label{fig:prob_S}}
\end{center}
\end{figure}

We show in \fref{fig:prob_S} the same probabilities as those shown in
\fref{fig:prob_L}, but the assumed value of $\theta_{13}$ is small ((a)
RSF-NOR-S; (b) RSF-INV-S).
\Fref{fig:prob_S}(a) indicates that the conversion probabilities from
the original $\nu_\mu$ are almost the same as those for RSF-NOR-L model;
as we have already noted, the value of $\theta_{13}$ does not matter in
the case of normal mass hierarchy.
On the other hand, the behaviours in \fref{fig:prob_S}(b) is
substantially different from those in \fref{fig:prob_L}(b).
In particular, the most easily detected flavour $\bar\nu_\rme$ is not
produced from the original $\nu_\rme$.
All these characteristics are consistent with the simple discussions
given in \sref{sec:Qualitative conversion schemes}.

\subsection{Neutrino signals at the Super-Kamiokande detector}
\label{sub:Neutrino signals at the Super-Kamiokande detector}

With the conversion probabilities given in the previous subsection and
the original neutrino spectrum by Thompson \etal \cite{Thompson03}, we
calculated the flux of each flavour neutrinos on the Earth.
(From this point on, we assume the galactic supernova neutrino burst and
take 10 kpc as a distance to the supernova.)
Using this flux and the cross section of the relevant neutrino
interaction at SK as well as the sensitivity of the detector, we can
calculate the expected event numbers from a future galactic supernova
neutrino burst.
In this paper, we adopt the most dominant reaction $\bar\nu_\rme {\rm p}
\to \rme^+{\rm n}$ alone; a cross section of the reaction has been
calculated in detail \cite{Vogel99}.

\begin{figure}[htbp]
\begin{center}
\includegraphics[width=15cm]{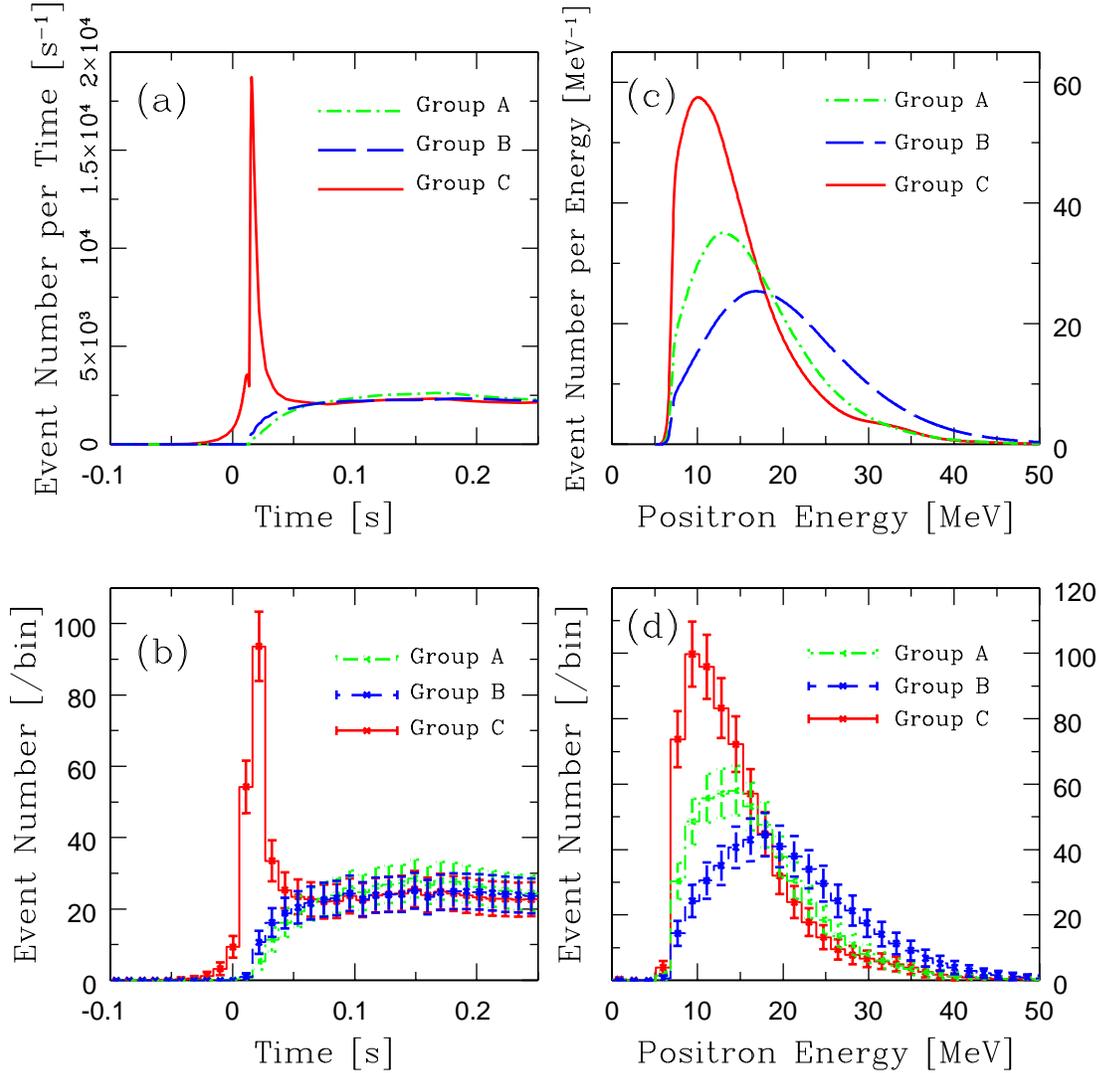}
\caption{Neutrino signal at the SK detector, which is evaluated for
 model groups A, B and C: (a) Time evolution of the neutrino signal in
 unit of [counts/s]; (b) the same as (a), but in unit of [counts/bin]
 with $1\sigma$ statistical error bars; (c) number spectrum of positrons
 for first 0.25 s after core bounce in unit of [counts/MeV]; (d) the
 same as (c), but in unit of [counts/bin] with $1\sigma$ statistical
 error bars. \label{fig:W95S_SK}}
\end{center}
\end{figure}

\Fref{fig:W95S_SK}(a) shows time evolution of the energy-integrated
event as a function of time for each group A, B and C, and we show the
same in \fref{fig:W95S_SK}(b) but using equally spaced bins.
From the time evolution of neutrino events, we cannot discern the groups
A and B, because they show almost the same time profile.
For group C, however, since the original $\nu_\rme$ are converted into
$\bar\nu_\rme$, the time profile shows steep neutronization peak, and
the event number contained in this peak is expected to be statistically
significant as clearly seen in \fref{fig:W95S_SK}(b); the event number
included in the most prominent three bins is $\sim 180$.
If the neutronization peak were actually obtained, it strongly indicates
that the model group C would be favoured; since the group C contains only
one model, RSF-INV-L, a great number of problems concerning the neutrino
properties would be solved at the same time.
In that case, the neutrino would have the nonzero magnetic moment, the
mass hierarchy would be inverted, and the value of $\theta_{13}$ would
be large enough for MSW-H to be adiabatic.

The expected event number per unit energy range, which is integrated
during the first 0.25 s after core bounce, is shown in figures
\ref{fig:W95S_SK}(c) and \ref{fig:W95S_SK}(d), in units of counts/MeV
and counts/bin, respectively.
From these figures, the model group C gives the softest spectrum, while
B the hardest and A an intermediate one.
In addition to the time evolution of the neutrino events, the number
spectrum would provide useful information on the flavour conversion
mechanism.
Although the available data are restricted in order to avoid
uncertainties concerning shock wave propagation, the obtained data would
be statistically significant.
Using the spectrum, degeneracy between the group A and B is expected to
be broken.

\section{Discussion}
\label{sec:Discussion}

\subsection{How far can we probe the neutrino properties from the
  supernova neutrino observation?}
\label{sub:How far can we probe the neutrino properties from the
supernova neutrino observation?}

The expected neutrino signal at the SK detector has been investigated
thus far.
However, the mechanism of supernova explosions is quite unclear, since
all the reliable numerical simulations have not succeeded in pushing the
shock wave to penetrate the entire core.
There may be several unknown processes which we have omitted so far, and
the original neutrino spectrum as well as its luminosity curve are still
controversial.
Therefore, we cannot trust characteristics of the calculation by
Thompson \etal \cite{Thompson03} in detail.
Instead, we use rather simple quantities in order to discuss the
conversion mechanisms from the neutrino signals; this approach is
expected to considerably reduce the dependence on supernova models.

\begin{figure}[htbp]
\begin{center}
\includegraphics[width=10cm]{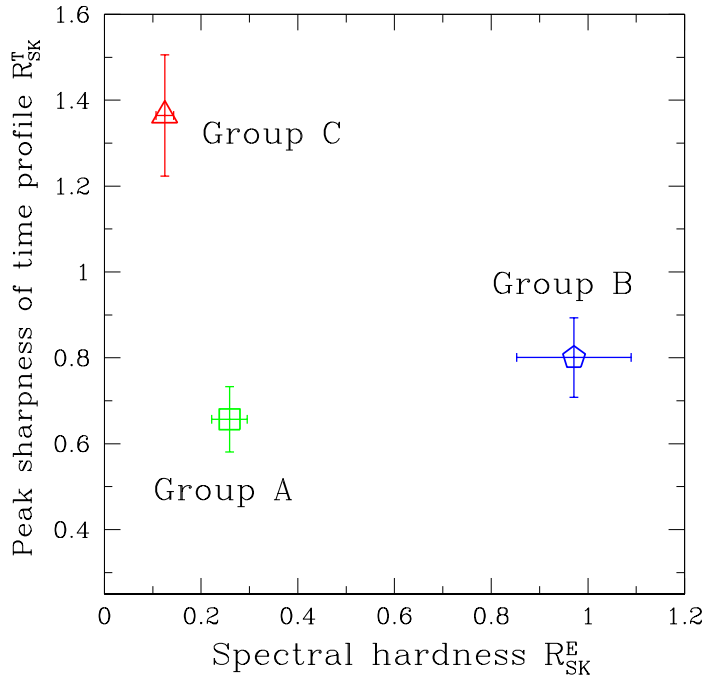}
\caption{The model groups A, B and C, plotted on $R_{\rm SK}^{\rm E}$
 vs $R_{\rm SK}^{\rm T}$ plane. The error bars include only statistical
 errors, and are at $1\sigma$ level. \label{fig:ratio_SK}}
\end{center}
\end{figure}

We adopt the following quantities:
\begin{eqnarray}
 R_{\rm SK}^{\rm E}&=&\frac{\mbox{Number of events for }E_\rme
  >25~{\rm MeV}}{\mbox{Number of events for }E_\rme <15~{\rm MeV}},
  \label{eq:R_SK E}\\
 R_{\rm SK}^{\rm T}&=&\frac{\mbox{Number of events for }0<t/{\rm ms}<75}
  {\mbox{Number of events for }75<t/{\rm ms}<150},
  \label{eq:R_SK t}
\end{eqnarray}
in order to represent the spectral hardness and the peak sharpness of
neutronization burst, respectively.
Since the time of the core collapse would never known with the neutrino
signal alone, we take time origin $t=0$ when the first neutrino signal
is detected.
The place of the model groups A, B and C on the ($R_{\rm SK}^{\rm E},
R_{\rm SK}^{\rm T}$) plane are shown in \fref{fig:ratio_SK}.
These groups are well separated from each other, and we expect that this
particular remains unchanged even if the adopted supernova model is
different.

\begin{figure}[htbp]
\begin{center}
\includegraphics[width=10cm]{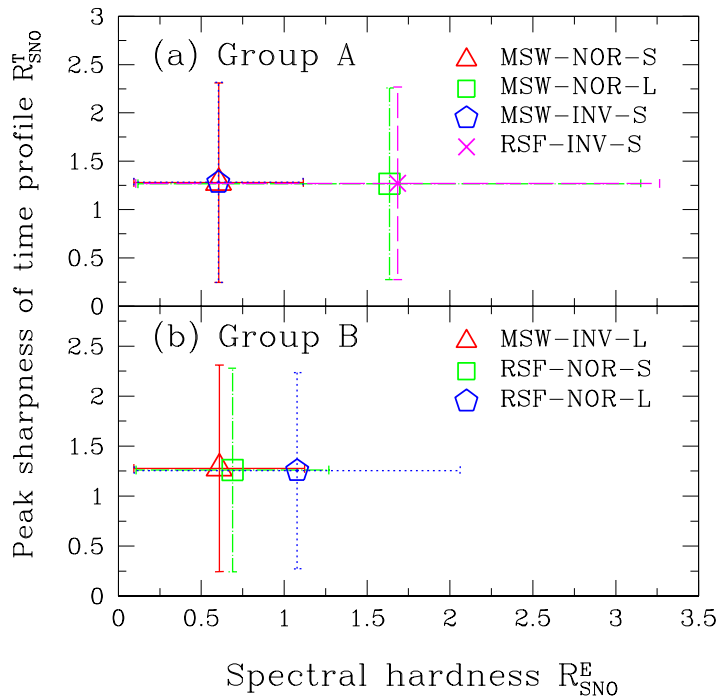}
\caption{The same as \fref{fig:ratio_SK}, but for the SNO detector. The
 models are plotted within (a) groups A and (b) B. \label{fig:ratio_SNO}}
\end{center}
\end{figure}

Although the degeneracy problem within each group cannot be solved by
the SK observation, which mainly detect $\bar\nu_\rme$, it may be
possible if $\nu_\rme$ could be detected efficiently.
SNO is such a detector currently data taking with 1 000 tons of heavy
water.
The supernova $\nu_\rme$ can be detected via $\nu_\rme\rmd\to e^-{\rm
pp}$ reaction.
Although the $\bar\nu_\rme$ are also detected through a similar
reaction, $\bar\nu_\rme\rmd\to e^+{\rm nn}$, these events could be
discriminated using delayed coincidence technique.
Thus at SNO, we expect that only $\nu_\rme$ signal can be extracted.
\Fref{fig:ratio_SNO} is the same as \fref{fig:ratio_SK} for the SNO
detector, but is plotted for the models in (a) group A and (b) B.
As shown in this figure, the degeneracy problem could be solved with the
$\nu_\rme$ data in principle, but the current smallness of the SNO
detector prevents a statistically significant discussion.

\subsection{Dependence on presupernova models}
\label{sub:Dependence on presupernova models}

Until this point, we have adopted the presupernova model W95S as our
reference model.
However, as already stated in \ref{sub:Presupernova profiles} or
investigated in \cite{Ando03d}, the RSF conversions are highly dependent
on the presupernova profiles.
In fact, figures \ref{fig:profiles} and \ref{fig:profiles_W02} show that
the relevant profile for the RSF effect, $\rho (1-2Y_\rme)$, is quite
different among these models as well as their metallicities.
On the other hand, the MSW conversions are insensitive to the
presupernova models because their relevant profile is $\rho Y_\rme$ that
is not subject to the deviation of $Y_\rme$ from 0.5.
In particular for the W02 models, the value of $\rho (1-Y_\rme)$
suddenly drops at the RSF-H region, yielding rather nonadiabatic
conversions.
Since RSF becomes incomplete but MSW does not change for these models,
the model group B, which contains RSF-NOR-S, RSF-NOR-L and MSW-INV-L
models, should be further divided into two subgroups; we define group B
I containing two RSF models, while MSW-INV-L is named group B II.
Consequently, we have four characteristic model groups A, B I, B II and
C; in particular, the groups B II and C contain only one specific model
each.
Within each group, we cannot identify each model from the SK
observation.

\begin{figure}[htbp]
\begin{center}
\includegraphics[width=15cm]{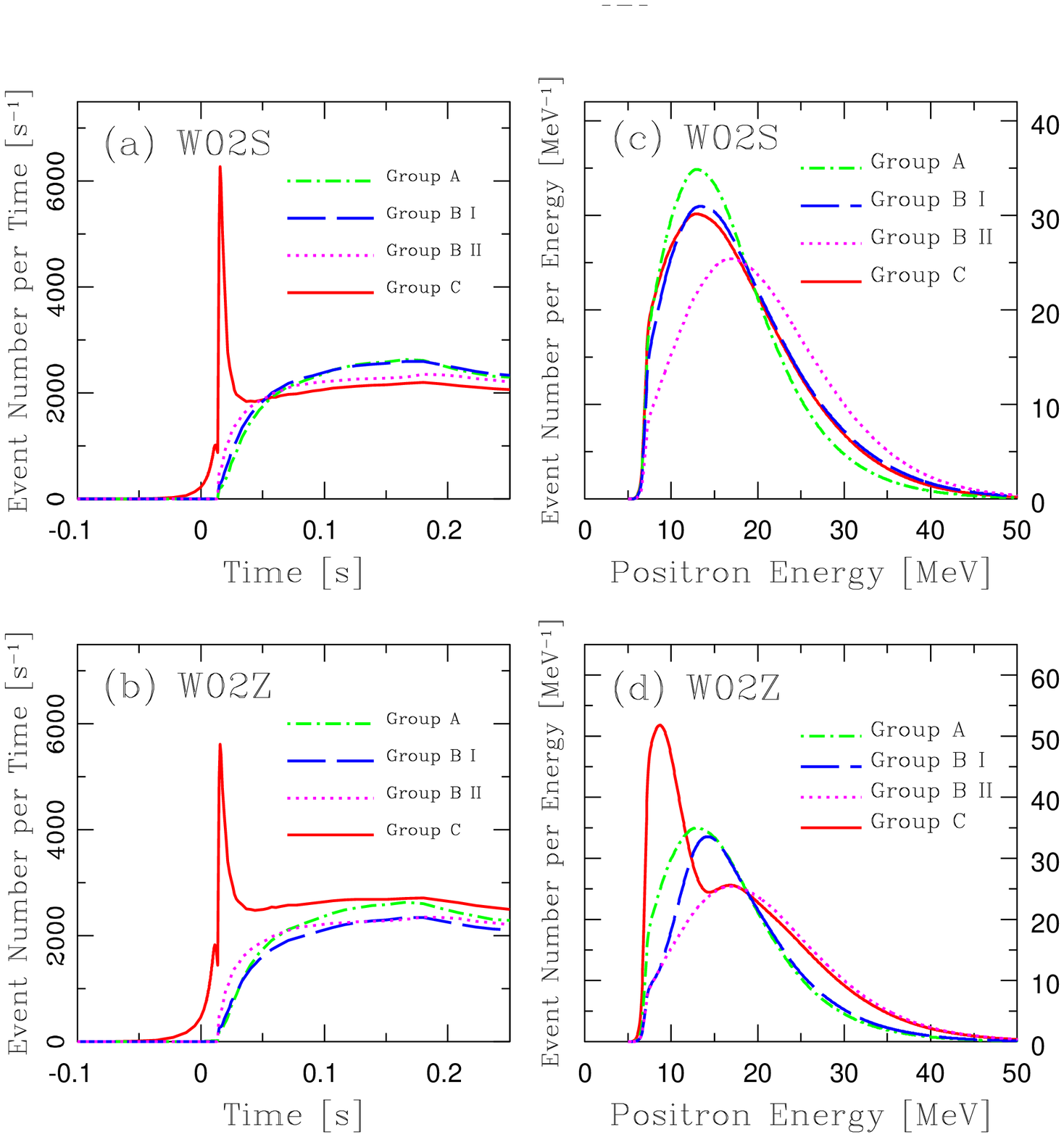}
\caption{Time evolution and number spectrum of neutrino events
 calculated with W02S and W02Z models. \label{fig:W02_SK}}
\end{center}
\end{figure}

Figures \ref{fig:W02_SK}(a) and (b) show time evolution of neutrino
events at SK, calculated with the W02S and W02Z models.
Comparing these figures with \fref{fig:W95S_SK}(a), the neutronization
peak is not as sharp as that for the W95S model.
This is because the RSF-H conversion is not efficient owing to sudden
drop of the $\rho (1-2Y_\rme)$ profiles.
However, since that peak is still prominent and statistically
significant, the model group C would be discriminated from the other
groups using the time profile detected at SK.

Figures \ref{fig:W02_SK}(c) and (d) show the number spectrum of the
neutrino signal, which is evaluated for the W02S and W02Z models,
respectively.
The spectra for groups A and B II are unchanged from the previous
calculation with the W95S model (\fref{fig:W95S_SK}(c)), because they
are essentially described by the pure MSW effect, which is insensitive to
presupernova models.
Therefore, we focus on the other two groups B I and C.
For the group B I, the hardness of spectra locates between that of group
A and B II, and it strongly depends on the adopted metallicities.
More detailed discussions concerning this group have been already given
in our previous paper \cite{Ando03d}, and we refer the reader to the
literature.
For the group C, the low energy peak disappears from the spectrum
calculated with the W02S model, reflecting the $\rho (1-2Y_\rme)$
profile of the W02S model (\fref{fig:profiles_W02}(a)), which
significantly reduces the efficiency of the RSF-H conversions.
On the other hand, this low energy peak remains for the W02Z model, also
reflecting the $\rho (1-Y_\rme)$ profile.
As clearly shown in \fref{fig:profiles_W02}(b), it mildly changes in the
RSF-H region for low energy neutrinos.
However, if the energy is beyond some critical value, RSF-H becomes
completely nonadiabatic, because the $\rho (1-2Y_\rme)$ profile abruptly
drops to zero.
Thus, the $\nu_\rme\to\bar\nu_\rme$ conversion is highly efficient for
low energy neutrinos but is highly inefficient for high energy ones; for
the high energy region the $\nu_{\rm x}\to\bar\nu_\rme$ conversion is
relevant, which leads to double peak profile of the spectrum.

\section{Conclusion}
\label{sec:Conclusion}

In this paper, we investigated the RSF conversions in supernovae for
both the normal and inverted mass hierarchy.
As the case for the pure MSW effect, we found that the RSF transitions
are strongly dependent on the neutrino mass hierarchy, and also on the
value of $\theta_{13}$.
We first gave qualitative discussion on the neutrino conversions
including both the RSF and MSW effect, for eight parameter sets
summarized in \tref{table:models}.
From that consideration, it was found that these models are categorized
into only three groups, each of which is expected to show characteristic
neutrino signal at the SK detector; we named these groups A, B and C,
and which parameter set is included in each group is summarized in
\tref{table:conversions}.

We, then, presented results of numerical calculations of flavour
conversions in supernova envelope.
The density and $Y_\rme$ profiles of the presupernova star calculated by
Woosley and Weaver \cite{Woosley95} (W95S) was adopted for the
calculations.
As the magnetic field structure, we assumed dipole-type and normalized
its strength at the surface of the iron core.
Using the conversion probabilities calculated by such a procedure and
the original supernova neutrino spectrum as well as luminosity curve
given by Thompson \etal \cite{Thompson03}, the expected neutrino signal
at the SK detector was estimated.
As the result, it was found that there are clear difference between the
model groups both for the spectral shape and the time evolution of the
neutrino events.
In particular, the model group C, which include the RSF-INV-L model
alone, shows a sharp neutronization peak.
Therefore, if this peak were detected in reality from the future
galactic supernova neutrino burst, it would strongly support the
RSF-INV-L model.
It would indicate that many problems concerning the neutrino property
should be solved at the same time, i.e., the neutrino have the nonzero
magnetic moment, the mass hierarchy is inverted, and the value of
$\theta_{13}$ is large enough to induce the adiabatic MSW-H resonance.
From the spectral shape, it would be possible to discern the groups A
and B, however, within each model group we need other information such
as the signal of $\nu_\rme$, to discriminate one model from the others.
Although the SNO detector could detect $\nu_\rme$ efficiently, current
smallness of the detector would prevent the statistically significant
discussions.

We also studied the dependence on presupernova models.
As already investigated in our previous paper \cite{Ando03d}, the RSF
conversion is strongly dependent on the deviation of $Y_\rme$ from 0.5,
which is quite sensitive to the metallicities as well as the weak
interaction rates adopted in the simulation of stellar evolution.
The presupernova models by Woosley \etal \cite{Woosley02} (W02S and
W02Z) shows the considerably different $\rho (1-2Y_\rme)$ profile from
that of the W95S model.
For both models, there exists sudden drop of the $\rho (1-2Y_\rme)$
profile at the RSF-H region, which pushes the RSF conversion rather
nonadiabatic, and the expected neutrino signal was found to be different
from that estimated with the W95S model.
Still, since the neutronization peak for the RSF-INV-L model exists at
the statistically significant level, it would keep to be useful method
also in these cases.

\ack
SA's work is supported by Grant-in-Aid for JSPS Fellows.
KS's work is supported in part by Grant-in-Aid for Scientific
Research provided by the Ministry of Education, Culture, Sports, Science
and Technology of Japan through Research Grant No S14102004,
Grant-in-Aid for Scientific Research on priority Areas
No 14079202.

\section*{References}
\bibliography{refs}

\end{document}